\documentstyle[preprint,aps,epsfig]{revtex}
\tighten
\def\bea{\begin{eqnarray}}
\def\beann{\begin{eqnarray*}}
\def\beq{\begin{equation}}
\def\eea{\end{eqnarray}}
\def\eeann{\end{eqnarray*}}
\def\eeq{\end{equation}}
\def\nn{\nonumber}

\def\ran{\rangle}
\def\lan{\langle}

\newcommand{\bcdot}{\bbox{\cdot}}

\newcommand{\bsigma}{\bbox{\sigma}}

\newcommand{\btau}{\bbox{\tau}}

\begin{document}
\headheight1.2cm
\headsep1.2cm
\baselineskip=20pt plus 1pt minus 1pt
%
%
\draft
\title{Chiral aspects of hadron structure}
\author{
A.W. Thomas$^{1}$ and  G. Krein$^2$ \\
{\small $^1$ Department of Physics and Mathematical Physics and Special 
Research Center for} \\
{\small the Subatomic Structure of Matter, University of Adelaide, SA 5005, 
Australia} \\
{\small  $^2$ Instituto de F\'{\i}sica Te\'{o}rica, Universidade Estadual 
Paulista}\\
{\small Rua Pamplona, 145 - 01405-900 S\~{a}o Paulo, SP, Brazil}\\
}
\maketitle
\begin{abstract}
Chiral loop corrections for hadronic properties are considered in a 
constituent quark model. It is emphasized that the correct implementation
of such corrections requires a sum over intermediate hadronic states.
The leading non-analytic corrections are very important for baryon
magnetic moments and explain the failure of the sum rule 
$(\mu_{\Sigma^{+}}+2\mu_{\Sigma^{-}})/\mu_{\Lambda}=-1$ predicted 
by the constituent quark model.
\end{abstract}
\vspace{2.5cm}
\noindent{PACS NUMBERS: 11.30.Rd, 12.39.Jh, 12.39.Fe, 12.40.Yx,13.40.E}

\vspace{1.0cm}
\noindent{KEYWORDS: Chiral symmetry, quark model, potential models, 
hadron spectrum, magnetic moments}

\newpage 

The role of chiral symmetry in hadron structure, spectroscopy and
hadron-hadron interactions is a recurrent theme in modern strong
interaction physics. Even lattice QCD cannot avoid the issue. Current
lattice simulations for light current quark masses are 
computationally intensive and present computer limitations mean that
lattice simulations are restricted to relatively large quark masses.
In order to make contact with the physical world 
extrapolation schemes must be devised. Guidance from quark 
models which are able to interpolate between the correct chiral and heavy 
quark limits of QCD is of importance for such extrapolations.

Extrapolations of lattice data 
which respect both the chiral behavior of QCD and the
heavy quark limit have recently been developed for baryon
masses~\cite{masses} and magnetic moments~\cite{magmom}.
The cloudy bag model~\cite{CBM} (CBM) proved extremely useful as a framework for
exploring such problems. As the CBM is based upon the MIT
bag as a model for the underlying quark structure, it would clearly be 
desirable to carry out similar studies with alternative models.
However, the essential feature of the CBM, which {\em must} be retained
in any treatment of hadron structure or hadronic interactions in order
to be consistent with the chiral structure of QCD, is that in
calculating chiral loops one must project onto intermediate {\em
hadronic} states~\cite{prev}. One must not calculate chiral loops at the
quark level, independent of the hadronic environment in which the quark
is found.

This particular point is of importance for the problem recently raised by 
Lipkin~\cite{Lipkin} concerning the sum rule for the ratio of the 
$\Sigma^{\pm}$ and $\Lambda$ hyperons     
\beq
R_{\Sigma/\Lambda} \equiv \frac{\mu_{\Sigma^+}+2\mu_{\Sigma^-}} 
{\mu_{\Lambda}} .
\label{Ratio}
\eeq
The quark model prediction is $R^{QM}_{\Sigma,\Lambda}=-1$. Using the 
experimental values from hyperon magnetic moments, one obtains 
$R^{Exp}_{\Sigma,\Lambda}= -0.23$. In the following we show that the 
leading non-analytic (LNA) chiral corrections 
are large and explain why the sum rule fails. Moreover, we explain why
the correct model independent LNA behavior is not obtained from chiral
loops on single quarks.

We start by reviewing the basic equations and concentrate on pionic corrections;
the extension to include kaon corrections is straightforward. The Hamiltonian 
of a chiral quark model can quite generally be written as
\beq
H = H_0 + H_{\pi} + W
\label{Hamilt}
\eeq
where $H_0$ describes the bare quark states $|B^{(0)}_{\alpha}\ran$ 
of the system
\beq
H_0 |B^{(0)}_{\alpha}\ran = E^{(0)}_{\alpha}|B^{(0)}_{\alpha}\ran .
\label{bare}
\eeq
In this, $\alpha$ represents the set of spatial, spin and isospin quantum 
numbers of the baryons. $H_{\pi}$~is the Hamiltonian for non-interacting 
pions and $W$ is the pion-quark interaction vertex. Note that in principle 
$H_0$ might contain not only the confinement interaction, but also hyperfine 
interactions, such as one-gluon exchange. 

Pionic corrections are calculated by projecting the Hamiltonian onto the 
single-baryon states $|B^{(0)}\ran \equiv B^{\dag}|0\ran$, where $B^{(0)\dag}$
is the creation operator of the bare baryon and $|0\ran$ is the vacuum state. 
The resulting effective baryon-pion Hamiltonian can be written schematically 
as
\beq
H = \sum_{\alpha} E^{(0)}_{\alpha} B^{\dag}_{\alpha} B_{\alpha} + 
\sum_j \omega_j \, a^{\dagger}_ja_j +
\sum_{j \alpha\alpha'}W^j_{\alpha\alpha'}B^{\dag}_{\alpha'} B_{\alpha} a_j 
+ {\rm h.c.}
\eeq
where $a^{\dagger}_j$ and $a_j$ are pion creation and annihilation operators 
and $j$ indicates isospin and spatial variables. (Note that we have
taken bare baryon states with different $\alpha$ indices to be 
orthogonal.)

The physical baryon mass $M_B$ (where $B$ stands for e.g. $N$, $\Delta$,
$\Sigma$, $\Lambda$ etc. ) can be computed by dressing the bare (quark
model) baryon with its meson cloud in a straightforward way:
\beq
M_B = M^{(0)}_B + \Sigma(M_B) 
\label{M_B}
\eeq
%
where $M^{(0)}_B$ is the bare baryon mass (i.e. without pionic corrections)
and the self-energy, $\Sigma(E)$, is given as
\bea
\Sigma(E) = \langle B_0|W^{\dagger}\frac{1}{E - \tilde H_0}W|B_0\rangle . 
\label{Sigma}
\eea
%
Here $W$ is the effective pion-nucleon vertex and $\tilde H_0$ is the 
single-particle Hamiltonian, evaluated with the {\em physical} masses of
the baryons and the pion~\cite{CBM}~\footnote{One can also
systematically choose the level of sophistication of the underlying quark
model by subdividing the space of bare hadron states into a $P$-space
which is dealt with explicitly and a $Q$-space whose effects are
parametrized~\cite{CBM}.}. Insertion of a sum over intermediate
baryon-pion states in Eq.~(\ref{Sigma}) leads to
\beq
\Sigma(E) = \sum_n \langle B_0|W^{\dagger}|n\rangle \frac{1}{E - E_n}
\langle n|W|B_0\rangle .
\label{sum}
\eeq

The structure vertex-propagator-vertex $W^{\dagger}(E - \tilde H_0)^{-1}W$ in 
Eq.~(\ref{Sigma}) is an effective quark-quark interaction. For a 
quark-quark interaction mediated by pion exchange, it is crucial that it
includes processes where the pion is emitted and re-absorbed by the same
quark, as well as those where it is exchanged between a pair of quarks.
As explained in Ref.~\cite{prev}, any treatment that
misses diagrams where the pion is 
emitted and absorbed by the same quark line in the hadron leads to 
incorrect conclusions concerning hadron properties. Such diagrams are 
essential for the correct spin-isospin dependence of the corrections and, in 
particular, to yield the correct leading non-analytic contributions~(LNA). 
On the other hand, in order to obtain the correct 
LNA corrections to hadron masses 
{\em it is not sufficient} to consider just loop diagrams on a quark line, 
independent of its environment.

Let us be more specific. Consider an interaction of the form (we restrict 
ourselves to the SU(2) case, but the argument is general): 
\beq
\sum_{i < j} \bsigma_i\bcdot\bsigma_j
\btau_i\bcdot\btau_j \,V_{ij} ,
\label{ilessj}
\eeq
where $V_{ij}$ is the radial part, not restricted to be a contact interaction.
Such an interaction is the basis of the calculations in several quark models,
in particular in the model of Glozman and Riska~\cite{GRrev}. The 
overall strength (in hadron $|H\rangle$) from the interaction of 
Eq.~(\ref{ilessj}) is given by the spin-isospin ($SI$) matrix element
\beq
\langle SI \rangle_{H} = \langle H |\sum_{i < j} \bsigma_i\bcdot\bsigma_j 
\btau_i\bcdot\btau_j |H\rangle ,
\label{SI}
\eeq
%
which yields, for the $N$ and the $\Delta$, $\langle SI \rangle_N = 30$ and 
$\langle SI \rangle_\Delta = 6$. These lead to the relations
\bea
M_N &=& M_0 - 15 P^{\pi}_{00} \label{delMN} \\
M_{\Delta} &=& M_0 - 3 P^{\pi}_{00} \label{delMD},
\eea
%
where $M_0$ is the corresponding unperturbed energy and $P^{\pi}_{00}
\simeq 30$~MeV is the fitting parameter corresponding to the radial matrix 
element of Eq.~(\ref{ilessj}), in the lowest-energy unperturbed shell of 
the 3-quark system. 

On the other hand, the correct field theoretic self-energy calculation leads
to
\bea
M_N &=& M_0 - \frac{25}{2} P^{\pi}_{00} - 16 P^{\pi}_{N\Delta}\label{MN1} \\
M_{\Delta} &=& M_0 - 4 P^{\pi}_{\Delta N} - \frac{25}{2} 
P^{\pi}_{00}\label{MD1},
\eea
%
where $P^{\pi}_{N\Delta}$ and $P^{\pi}_{\Delta N}$ differ from $P^{\pi}_{00}$ 
by a factor $\Delta M = M_{\Delta}-M_N$ in the energy denominator 
(see Eq.~(\ref{sum}) and Fig.~1). 
Note that the form of these equations is general, in the
sense that the chiral limit does not alter this structure. Also, it should be 
noted that because of the dependence on $\Delta M = M_{\Delta}-M_N$ of the 
self-energies $P^{\pi}_{N\Delta}$ and $P^{\pi}_{\Delta N}$, these equations 
go beyond simple perturbation theory (recall the difference between 
Rayleigh-Schr\"odinger and Wigner-Brillouin methods). 

\vspace{1.0cm}
\begin{center}
\epsfig{angle=0,figure=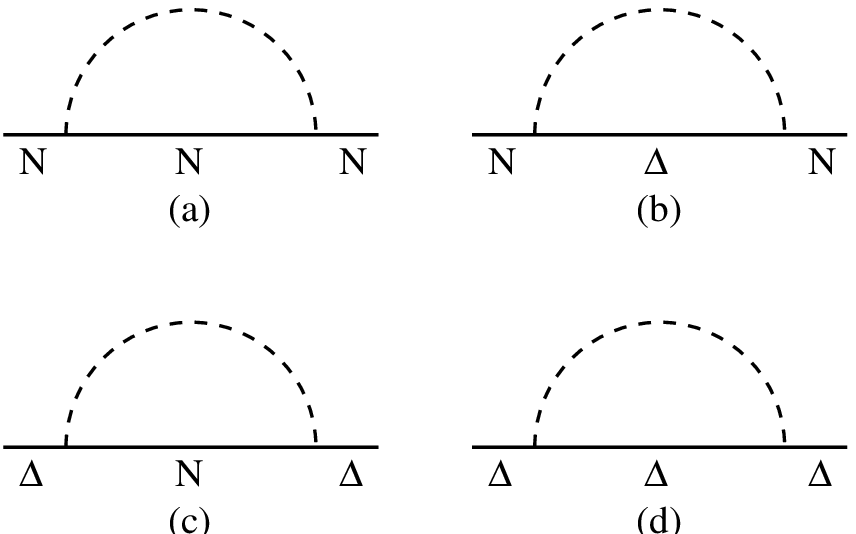}

\vspace{0.5cm}FIGURE 1. One-loop pion self-energy corrections to the
nucleon
($N$) and delta ($\Delta$).
\end{center}

\vspace{1.0cm}

Suppose for the moment that $P^{\pi}_{N\Delta} = P^{\pi}_{\Delta N}$. Then
we find
\beq
M_{\Delta} - M_N = 12 P^{\pi}_{N\Delta} .
\eeq
\noindent
Of course, $P^{\pi}_{N\Delta}$ can now be fitted to the experimental value of
$M_{\Delta}-M_N$, and nothing apparently changes with respect to the 
result of Eqs.~(\ref{delMN}) and (\ref{delMD}). But this is not the entire 
story, since this then implies a huge nucleon self energy. To estimate this, 
suppose $P^{\pi}_{N\Delta} = P^{\pi}_{00}$, which is equivalent to setting 
the  $\Delta$-$N$ mass difference to zero in the radial integrals. Then, the 
nucleon self-energy would be given by $ -57/2 P^{\pi}_{00}$, instead of 
$ - 15 P^{\pi}_{00}$. This would imply a total 
nucleon self-energy of -$855$~MeV;
a pretty big self-energy indeed! The situation becomes even worse in practice,
because the $\Delta$-$N$ mass difference is quite large and therefore 
$P^{\pi}_{00} > P^{\pi}_{N\Delta}$. Of course, it is very hard to justify 
nucleon self-energies of the order of the nucleon rest mass in a 
non-relativistic framework. In addition, in the process of calculating 
self-energy corrections for the low-lying and excited states, the incorrect 
treatment of intermediate hadronic states will lead to fatally incorrect 
systematics.

Let us now consider the problem raised by Lipkin~\cite{Lipkin}. 
The magnetic moment of a baryon $B$ is defined by $\mu_B = G_M^B(0)$, 
where $G_M^B (q^2)$ is the magnetic form factor. In Fig.~1 we show the
different contributions to $G_M^B (q^2)$. The calculation of $G_M^B (q^2)$
proceeds on the same lines as for the cloudy bag model~\cite{CBM};
the explicit expressions (including both octet and decuplet baryons in
the intermediate states) were derived in Ref.~\cite{TheMiTho}. The leading 
non-analytic contributions to $\mu_B$ come from intermediate states with the 
same quantum numbers as the external ones, i.e. from the terms
$C = B$ in Fig.~1(c). The explicit form of this contribution is
\beq
G^{1(c)}_M(0)= 2m_N \frac{1}{16\pi^2 f^2_{\pi}}\,\beta^{BB\pi}
\int dk \, \frac{k^4 u^2(k)}{w^4(k)}
\label{G1c}
\eeq
where the $\beta^{BB\pi}$ are given by SU(3). For the case of interest
here they are given by
\bea
&&\beta^{NN\pi} = (F+D)^2\,\lan N|\tau_3|N\ran \nn\\
&&\beta^{\Sigma^-\Sigma^-\pi} = - \beta^{\Sigma^+\Sigma^+\pi} =
2\left(\frac{1}{3}D^2 + F^2 \right) \nn\\
&&\beta^{\Lambda\Lambda\pi} = 0
\label{betas}
\eea
where $F$ and $D$ are usual SU(3) axial couplings.
\vspace{1.0cm}
\begin{center}
\epsfxsize=13.cm\epsfbox{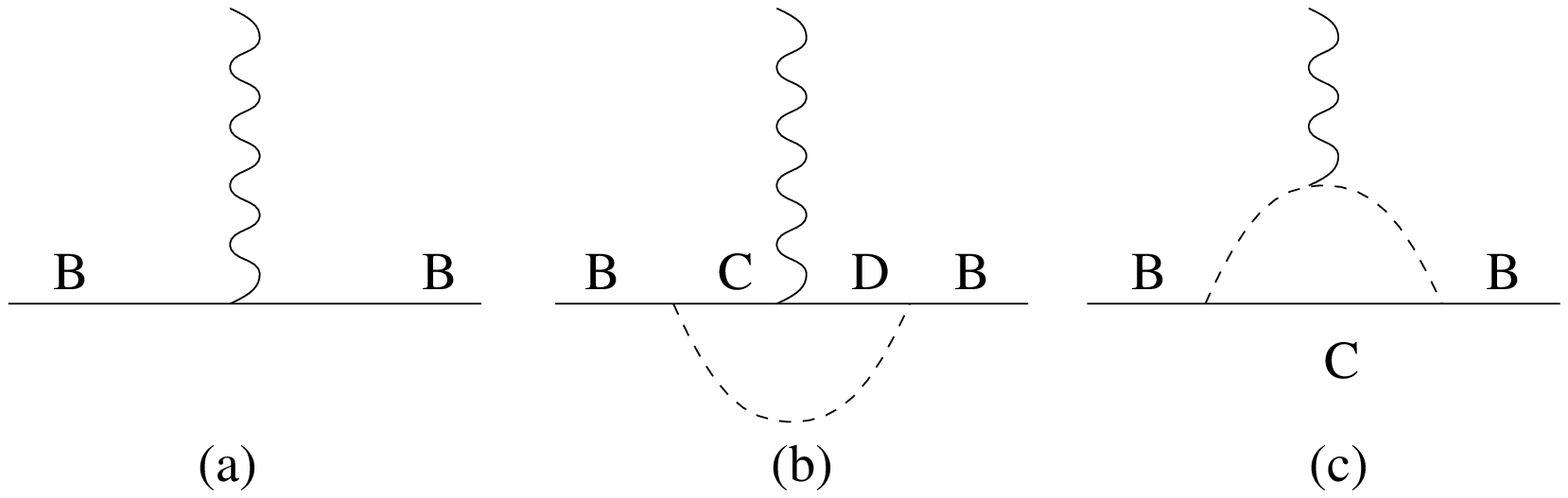}
 
\vspace{0.5cm}FIGURE 1. The various contributions to the magnetic moment of a 
baryon.
\end{center}

The LNA contribution for $G_M(0)$ is easily obtained from Eq.~(\ref{G1c}) and 
is given by 
\beq
\mu^{LNA}_B = \frac{m_N m_\pi}{8\pi f^2_{\pi}}\,\beta^{BB\pi}
\equiv \alpha_B \, m_\pi .
\label{LNAmm}
\eeq
Note that this is precisely what is obtained in Refs.~\cite{ChiPT1}~
\cite{ChiPT2} with chiral perturbation theory -- as it must, because
both calculations are based upon chiral symmetry.
The LNA behavior does not necessarily dominate the physics of
hadronic properties in general. However, in
the case of the magnetic moments the fact that the LNA term is also the
first term in a power series for $\mu_B$ as a function of $m_\pi$ and
that the numerical value of the LNA coefficient is large means that it
is phenomenologically important there. Over the range $m_\pi \, \epsilon \,
(0,2 m_\pi^{\rm phys})$ it is quite a good approximation to write the
baryon magnetic moments as:
\beq
\mu_B = \mu_B^0 + \alpha_B \, m_\pi + {\cal O}(m_\pi^2) .
\label{MMexp}
\eeq
Applying this to the $\Sigma^{\pm}$ and $\Lambda$ magnetic moments, with
the one loop chiral coefficients of Ref.\cite{ChiPT1} one finds:
\begin{eqnarray}
\mu_{\Sigma^+} & \simeq & 2.80 - 2.46 m_\pi \label{sig+} \\
\mu_{\Sigma^-} & \simeq & -1.50 + 2.46 m_\pi \label{sig-} 
\end{eqnarray}
with $m_\pi$ in GeV. The $\Lambda$ magnetic moment is just equal to the
experimentally measured moment up to order $m_\pi^2$. We stress that the
coefficient of the $m_\pi$ term is model independent and that the 
non-analytic terms are large -- for example, 
roughly $\frac{1}{3}$ of the measured $\Sigma^-$ magnetic moment at the
physical pion mass.

Using these expressions we can now evaluate the ratio of magnetic
moments considered by Lipkin, as a function of the pion mass:
\beq
\frac{\mu_{\Sigma^+}+2\mu_{\Sigma^-}} {\mu_{\Lambda}}
\simeq 0.33 - 0.56 \frac{m_\pi}{m_\pi^{\rm phys}} + {\cal O}(m_\pi^2).
\label{lipR}
\eeq
This expression makes it exceptionally clear why the sum-rule fails. It
is extremely sensitive to the value of the pion mass because it involves
a LNA piece of the $\Sigma$ magnetic moment that has not been arranged
to cancel in some way. While the ratio is only -0.23 at the physical pion
mass, way below the naive expectation of -1, and in the chiral limit it
even has the opposite sign (+0.33), at just above twice the physical
pion mass it would take the expected value. We note that such behavior
could never be reproduced within a constituent quark model, where the
constituent mass would vary by a mere 10-20 MeV as $m_\pi$ varies over
$(0, 2 m_\pi^{\rm phys})$.

Next we examine whether the LNA contributions can be equivalently
included in the constituent quark mass~\cite{GLrep}. The essence of chiral
perturbation theory, as a phenomenological implementation of the chiral
symmetry of QCD, is that there are certain non-analytic terms in the
exact expression for any hadronic property, as a function of the 
current quark masses, which are {\em model independent}. These terms are
determined by just a few gross hadron properties, including their axial
charges. For the nucleon mass the leading non-analytic (LNA) term is:
\beq
\delta M_N = - \frac{3}{32 \pi f_\pi^2} g_A^2 m_\pi^3.
\label{LNA_N}
\eeq
This comes from the loop diagram in Fig. 1(a), with the pion-nucleon
coupling given by PCAC as $\frac{g_A}{2 f_\pi}$. In terms of quarks this
diagram necessarily involves pion exchange between {\em both} $i=j$ and
$i \neq j$, the essential point being that the 3-quark intermediate
state is a {\em nucleon} and hence degenerate with the initial baryon
state. 

In Ref.~\cite{GLrep} it is argued that the LNA contribution comes from
{\em quark} self-energy loops. It is emphasized that the non-relativistic 
quark model connection between $g_A$ (nucleon understood) and $g_A^q$ 
(the quark axial charge), namely $g_A^q = \frac{3}{5} g_A$, satisfies:
\beq
3 \left[ g_A^q \right]^2 = \frac{27}{25} g_A^2 \approx g_A^2.
\label{coinc}
\eeq
This argument, which attributes the discrepancy to the use 
of exact SU(6) is misleading. We note that Eq.~(\ref{LNA_N}) is {\em model
independent} -- it cannot depend on the model used to describe nucleon
structure. As an extreme example, we note that if the relativistic quark
model were used to relate $g_A^q$ and $g_A$ we would find:
\beq
3 \left[ g_A^q \right]^2 = \frac{27}{25 (0.65)^2} g_A^2 \approx 2.6
g_A^2,
\label{coinc2}
\eeq
where the factor of 0.65 is the well known relativistic correction for
massless quarks given by (for example) the MIT bag model \cite{MIT}.
The discrepancy in this case is now 260\%. 

To summarize, the 
suggestion that one can incorporate the chiral loop on the individual 
quark lines into the definition of the 
constituent quark mass is inappropriate
for many reasons. Firstly, if one were to use such a mass to define a Dirac
magnetic moment for the quark one would build in an incorrect LNA
contribution. Secondly, for an unstable particle like the $\Delta$ the
mass has an imaginary piece arising from the decay to $N \pi$. In order
to obtain the correct imaginary piece 
(i.e., the correct width of the $\Delta$)
one {\em must} include the imaginary part associated with the pion being
emitted and absorbed by the same quark -- an imaginary part that is
omitted in a naive calculation which ignores the environment in which
the constituent quark is sitting. Finally, if one were to compute chiral
corrections to the magnetic moments of the quarks themselves they would
be model dependent, for the reasons explained around Eqs.~(\ref{coinc})
and (\ref{coinc2}). This is in contradiction with QCD, within which the
LNA correction should be model independent. 

In conclusion we have shown that the LNA corrections to hadron properties 
have important practical consequences. In particular, they explain 
the reason for the failure of the sum-rule for hyperon magnetic moments
noted by Lipkin. We have also shown that the correct
LNA contributions to hadronic properties such as masses and magnetic
moments {\em cannot} be obtained by calculating loops at the quark
level, independent of the hadronic environment.

This work was supported in part by the Australian Research Council and
CNPq (Brazil).


\begin{thebibliography}{99}
%
\bibitem{masses} D. B. Leinweber, A. W. Thomas, K. Tsushima,
S. V. Wright, hep-lat/9906027, to appear in Physical Review D.
%
\bibitem{magmom}  Derek B. Leinweber, Ding H. Lu, Anthony W. Thomas,
Phys. Rev. D 60 (1999) 034014. 
%
\bibitem{CBM} S. Th\'eberge, A.W. Thomas and G.A. Miller, Phys. Rev. D 22 
(1980) 2838; Erratum-ibid. Phys. Rev. D 23 (1981) 2106; A.W. Thomas, 
Adv. Nucl. Phys. 13 (1984) 1; G.A. Miller, Int. Rev. Nucl. Phys. 2 (1984) 
190.
%
\bibitem{prev} A.W. Thomas and G. Krein, Phys. Lett. B 456 (1999) 5.
%
\bibitem{Lipkin} H. J. Lipkin, hep-ph/9911261.
%
\bibitem{GRrev} L. Ya. Glozman and D. O. Riska, Phys. Rep. 268 (1996) 263.
%
\bibitem{TheMiTho} S. Th\`eberge, G.A. Miller and A.W. Thomas, Can. J.
Phys. {\bf 60} (1982) 59.
%
\bibitem{ChiPT1} E. Jenkins, M. Luke, A. V. Manohar, and M. J. Savage,
Phys. Lett. B 302 (1993) 482.
%
\bibitem{ChiPT2} L. Durand and P. Ha, Phys. Rev. D 58 (1998) 013010.
%
\bibitem{GLrep} L. Ya. Glozman, Phys. Lett. B 459 (1999) 589.
%
\bibitem{MIT} T. de Grand, R.L. Jaffe, K. Johnson, and J. Kiskis, Phys.
Rev. D 12 (1975) 2060. 
%
\end{thebibliography}
\end{document}